\documentclass[runningheads]{llncs}
\newcommand\tab[1][2.7cm]{\hspace*{#1}}
\usepackage{graphicx}
\usepackage{float}
\begin{document}

\title{Multi-source Data Mining for e-Learning}
%
%
\author{Julie Bu Daher \and Armelle Brun \and Anne Boyer}
\authorrunning{J. Bu Daher et al.}
%
\institute{Universit\'e de Lorraine, Loria lab., Nancy, France (\url{http://kiwi.loria.fr})\\
\email{\{julie.bu-daher, armelle.brun, anne.boyer\}}@loria.fr}

\maketitle              

\section{Introduction} 
Data mining is the task of discovering interesting, unexpected or valuable structures in large datasets and transforming them into an understandable structure for further use \cite{chakrabarti2006data}. Different approaches in the domain of data mining have been proposed, among which pattern mining is the most important one. 
Pattern mining discovers various types of statistically relevant patterns, from which association rules can be generated. 

Recommender systems aim at identifying the items  that match a user profile (preferences, expectations, etc.), in order to recommend these items to the associated user. Pattern mining, and more specifically association rules mining, is one possible approach in recommender systems, where the items (or item-sets or sequences of items) that compose the consequence of the rules are those recommended. Rule-based recommender systems have a major advantage that lies in the fact that the recommendations they provide can be explained, through the use of the antecedent. This is an important characteristic in decision-based algorithms, especially when dedicated to human. 

Pattern mining has attracted much attention for many years, and with the emergence of  big data, it has become even more important. Some of the most common challenges include reducing the complexity of the process, avoiding the redundancy within the patterns, discovering important patterns, etc.\cite{aggarwal2014frequent}.

So far, pattern mining has focused mainly on the mining of one data source with a single data type. 
However, with the increase of the amount of data,  in terms of  volume, diversity of sources and nature of data, mining complex, multi-source, heterogeneous and multi-relational data is now an emerging challenge in the data mining community. This challenge is also the focus of our current work.




\section{Related Work} 
 
We give a brief overview of the types of data and associated algorithms in the literature, especially in the case of complex data. 
Complex data are data collections in which the data items may have different data types. 
It is now common that the data mined may not come from a unique source. Such kind of data is called {\bf multi-source data}. When there are relations between the sources or between the data dimensions, the data is called {\bf multi-relational} \cite{padhy2012multi}.



Two main approaches have been proposed to mine multi-source and multi-relational data \cite{wang2018review}.
The first approach manages all the data sources together in a unique process by combining all data sources together, through the use of the relations. Pinto et al., \cite{pinto2001multi}  introduce three algorithms that mine all the dimensions in a single process, and the resulting patterns contain information from all sources. Plantevit et al., \cite{plantevit2010mining} propose an approach which treats all the data together to consider different dimensions and levels of granularity at the same time. Egho et al., \cite{egho2012healthcare} combine information coming from different sources in one process and incorporates background knowledge in the form of hierarchies over data attributes. The other approach  mines  data sources separately and then combines the outputs. Hu et al., \cite{hu2007scalable} mine multi-dimensional sequences locally, in a distributed manner, where the resulting local patterns are grouped to form a unique result.

To summarize, current approaches in multi-source and multi-dimensional data mining either: (1) consider data sources independently and then merge the outputs or (2) combine data sources together in one source and manage a single-source output. The former approach may lead to a loss of information  and thus to a decrease in accuracy and coverage. The latter approach leads to highly complex algorithms, even if some works mine sources sequentially.


\section{Our vision and Planned Approach} 


Mining a single data source has a major advantage that lies in the low computational complexity of the mining process. However, we are convinced that both the accuracy of the associated model (the resulting set of patterns) and its coverage are not high due to the lack of information in a single data source. Having more than one data source, thus more information, results in higher accuracy and  coverage. 
Nevertheless, the complexity of the mining process depends on the number of data sources; the higher the number of sources, the higher the complexity of the mining process. 

In this work, we aim at designing a new mining algorithm that manages all the data sources and that has a limited complexity. We propose to exploit the links that exist between data sources or between elements of data sources, systematically or upon need,  for example when the process faces lack of information from some sources. More precisely, some links and sources may be exploited systematically, and others may be used only when required by the mining process, which is a way to reduce the complexity. The questions raised are thus: how to design an algorithm that selectively mines some sources, based on which criterion, what strategy to adopt, etc.? Is it only a matter of amount of information, or also of types of links and sources, or of computation time?

\begin{figure}[H]
 
   \includegraphics[width=0.5\textwidth, height=4cm]{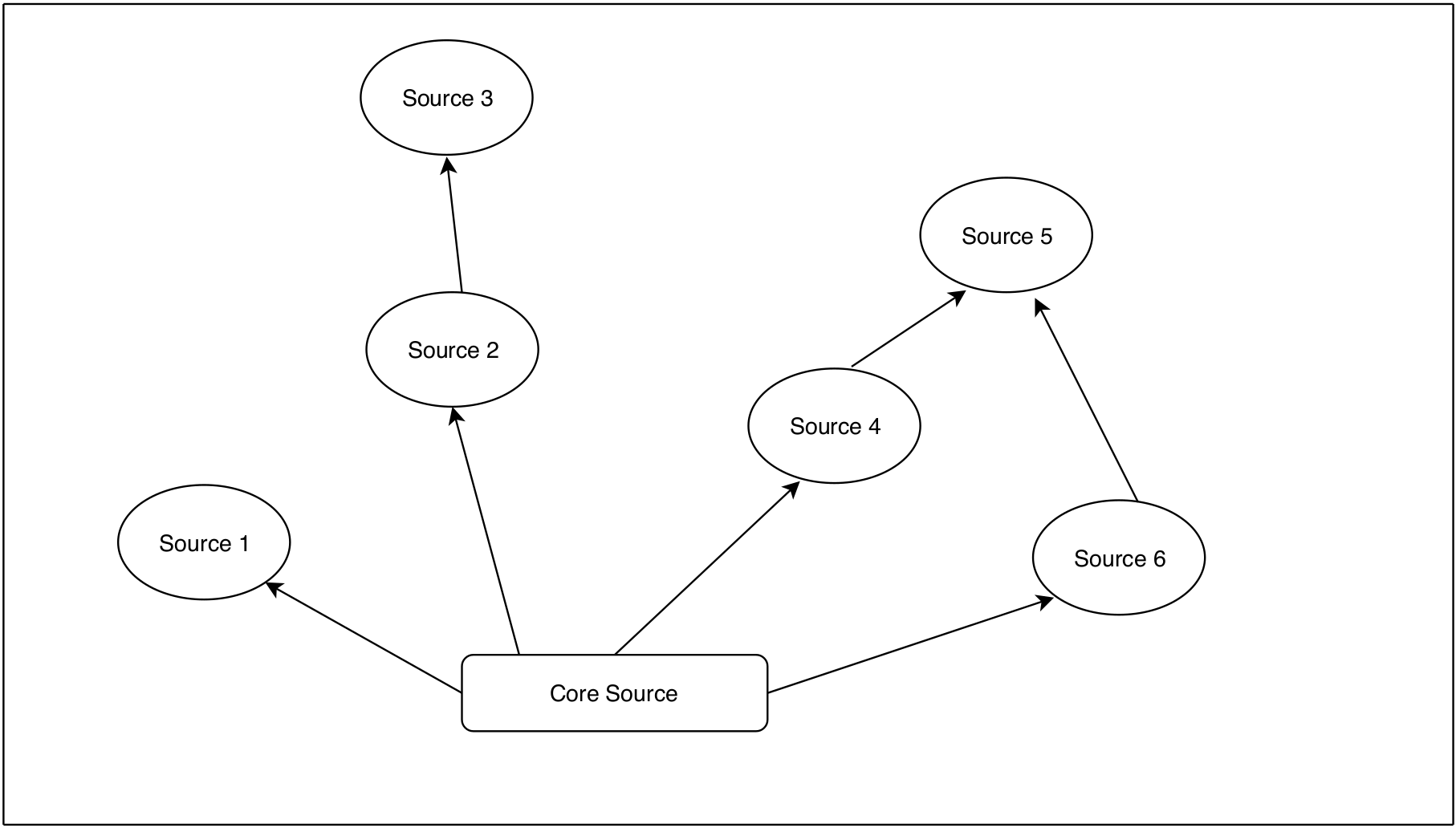} 
    \includegraphics[width=0.5\textwidth, height=4cm]{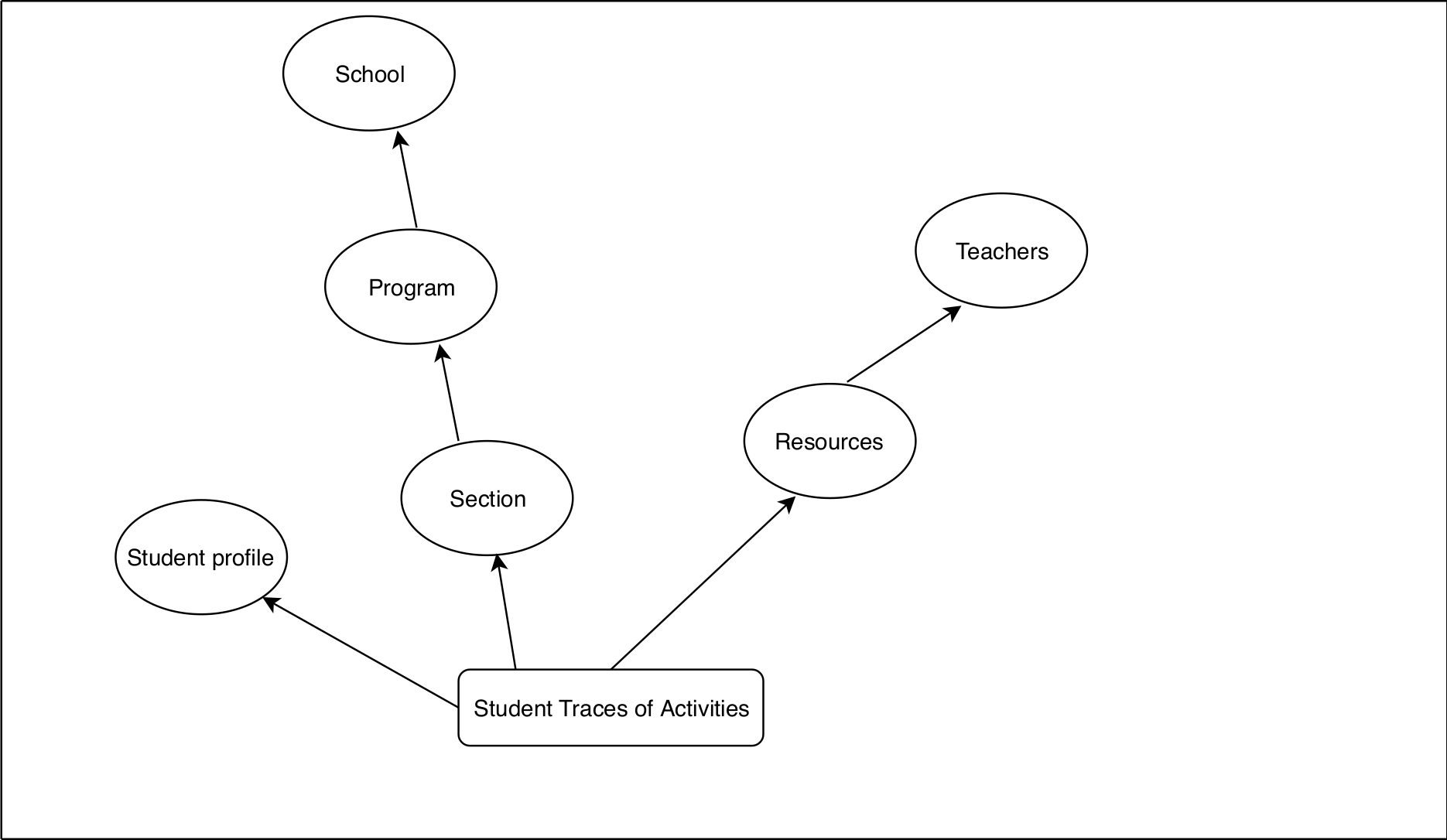}
     \tab (a) \tab ~~~~~~~~~~~~~~~~~~~~~~~~(b)
    \caption{(a) Overview of possible links and sources (b) On the METAL project}
    \label{fig: METAL}

\end{figure}

Concretely, we propose to view the data as follows. We consider one data source as the core source , and it consists of a set of sequences of events, for example user activities. The other data sources contain additional descriptive data about the elements of the events. The core source is systematically and extensively mined, while other sources are viewed as background information, or more general information. These additional sources can be linked to the core source or to other sources, and will be mined if needed.
Figure \ref{fig: METAL}~(a) represents a general view of the data sources and their relations.
This view is made up of several interlinked dimensions, forming a {\bf multi-dimensional} dataset.

\subsection{The METAL project}


We apply this work in the frame of METAL project that is in the domain of e-learning for primary and secondary school students in France. The general aim of the METAL project \footnote{http://metal.loria.fr} is to improve the quality of the learning process of school students. This improvement is attained through the development of an educational barometer intended for students where each student can know his/her detailed academic level and  will be provided with recommendations of pedagogic resources (exercises, lectures, exams, etc.) to increase his/her motivation and to improve his/her academic level. These recommendations will be provided based on their academic needs, preferences, past behavior on the virtual learning environment (VLE), profile, etc.
The recommendations are computed based on data about the students, teachers, programs, pedagogic resources, etc., each of them coming from a different source. This data is a real example of multi-source and multi-relational data, as described previously. Figure \ref{fig: METAL}(b) represents an overview of the data sources of METAL project.
We consider the "students' traces of activity" as the core data source. Other sources, like student profile, pedagogic resources, classes composition, teachers, school information, etc. are linked to the core data source, or to another data source. These other sources provide background information that is considered as more general information. 
The pattern mining algorithm will be designed to mine frequent sequential patterns of events, mainly from the core source, and from other sources if needed. The mining process will take into account the different sources and the relations between them. 

In the domain of e-learning, multi-source data is of high importance to perform high quality recommendations. As the sources provide background and more general information related to the activities of students, we expect to mine more frequent and more general patterns than the patterns mined when there is a single data source. Indeed, the data from additional sources  will be used to cope with the minimum support problem when mining patterns from a unique source. Furthermore, the additional information of each source could be at different levels of granularity which is useful in providing more general frequent patterns. This generalization results in higher data coverage that helps in providing more general results when specific patterns are not frequent. Here are some examples of frequent patterns that we intend to mine ($R_1$...$R_n$ are the $id's$ of the pedagogic resources that are included in the sequences of students' activities: 
$Rule_1: R_3\rightarrow\{R_8 R_{13} R_{27}\}~~~~$ 
$Rule_2:15 years~Metz~male~R_3 \rightarrow R_8~R_{13}~R_{27}$\\
$Rule_3: R_3~R_7~Math~R_{24}~\rightarrow~R_8~R_{27}$

\section{Conclusion}
The increase in the amount of data sources in data mining leads to several challenges, especially  related to the management of the numerous relations between sources, of various data types, and of the complexity of the mining task.
The challenge that we focus on is mining these heterogeneous and multi-relational data sources with limited complexity.

\section*{Acknowledgments}
This work has been funded by the PIA2 e-FRAN METAL Project.



 \bibliographystyle{splncs04}
\bibliography{mybiblio.bib}

\end{document}